\documentclass[conference]{IEEEtran}
\IEEEoverridecommandlockouts
\usepackage{cite}
\usepackage{amsmath}
\usepackage{amsfonts}
\usepackage{color}
\usepackage{amssymb}
\usepackage{algorithm}
\usepackage{algorithmic}
\usepackage{array}
\usepackage{url}
\usepackage{graphicx}
\usepackage{subfigure}
\usepackage{diagbox}
\usepackage{dsfont}
\usepackage{array}
\usepackage{bbding}
\usepackage{textcomp}
\usepackage{multirow}
\usepackage{threeparttable}

\usepackage[top=0.75in, bottom=1in, left=.625in, right=.625in]{geometry}

\IEEEoverridecommandlockouts
\begin{document}

\title{Online Task Scheduling for Fog Computing with Multi-Resource Fairness
\thanks{*The corresponding author Ziyu Shao is with School of Information Science and Technology, ShanghaiTech University, Shanghai 201210, China.}
}

\author{\IEEEauthorblockN{Simeng Bian, Xi Huang, Ziyu Shao*}
\IEEEauthorblockA{School of Information Science and Technology, ShanghaiTech University\\
Email: \{biansm, huangxi, shaozy\}@shanghaitech.edu.cn}
}

\maketitle

\begin{abstract}
In fog computing systems, one key challenge is \textit{online task scheduling}, \textit{i.e.}, to decide the resource allocation for tasks that are continuously generated from end devices.
The design is challenging because of various uncertainties manifested in fog computing systems; \textit{e.g.}, tasks' resource demands remain unknown before their actual arrivals.  
Recent works have applied deep reinforcement learning (DRL) techniques to conduct online task scheduling and improve various objectives.
However, they overlook the multi-resource fairness for different tasks, 
which is key to achieving fair resource sharing among tasks but in general non-trivial to achieve.
Thus it is still an open problem to design an online task scheduling scheme with multi-resource fairness.
In this paper, we address the above challenges.
Particularly, by leveraging DRL techniques and adopting the idea of dominant resource fairness (DRF),
we propose \textit{FairTS}, an online task scheduling scheme that learns directly from experience to effectively shorten average task slowdown while ensuring multi-resource fairness among tasks. 
Simulation results show that FairTS outperforms state-of-the-art schemes with an ultra-low task slowdown and better resource fairness.
\end{abstract}

\section{Introduction} 
To accommodate ever-increasing computation demand and the diversity of real-time services, 
in recent years, there has been a growing trend in extending computation capacities of cloud to the edge of the network.
In this process, fog computing has emerged as the most promising paradigm to deliver best quality-of-service\cite{mukherjee2018survey}.

In typical fog computing systems, one of the main challenges is \textit{online task scheduling}, \textit{i.e.}, how to allocate available resources to newly arriving tasks that are generated by end devices in an online fashion. 
In general, designing an effective online task scheduling scheme is challenging, 
because fog computing systems often manifest various uncertainties; 
for example, task features such as its resource demand are often unknown before its actual arrival. 
To cope with such uncertainties, an effective scheme must be able to \textit{learn} from timely feedback incurred by its prior decisions 
{to} \textit{guide} its subsequent decision making process.

Deep reinforcement learning (DRL) technique has been applied in many areas of artificial intelligence and has achieved many encouraging early results. The key advantage of this technique is that it requires no prior knowledge of the system, and can be adapted to various objectives.
Notably, recent works have actively applied DRL to solve related online task scheduling problems. 
For instance,  
Liu \textit{et. al.} \cite{liu2019resource} proposed an effective task scheduling scheme that achieves a trade-off between power consumption and task latency.
Ye \textit{et. al.}  \cite{ye2018deep} developed a decentralized
resource allocation mechanism for vehicle-to-vehicle (V2V) systems to improve latency and reliability. 
Zhang \textit{et. al.} \cite{zhang2018deep} applied DRL to solve the task scheduling problem and designed a DQN-based policy for mobile users to minimize its monetary cost and energy consumption.

For those online task scheduling schemes, despite various performance improvements they have achieved, they generally ignore the resource fairness among tasks, which is another critical objective to attain in the process of task scheduling. 
Unfair resource allocation will degrade the willingness of IoT users to use the fog system.
Basically, a fair task scheduling possesses various desirable attributes\cite{ghodsi2011dominant}.
For example, it ensures a balanced resource sharing among tasks, so that no tasks would be assigned with excessive resources which results in performance degradation of any other tasks. 
Besides, resource fairness encourages to maximize the utilization of available resources, which conduces to more effective resource management. 

So far, it still remains as an open and challenging problem 
to design an online task scheduling scheme with resource fairness.
On one hand, in a typical fog computing system, task scheduling usually involves multiple types of resources \cite{mukherjee2018survey}. 
Meanwhile, tasks from different end devices also vary widely in their resource demands.
Given task demand and resource type variety, it has been proven non-trivial to achieve \textit{multi-resource fairness} \cite{ghodsi2011dominant}, not to mention to incorporate it into the online task scheduling process.
On the other hand, a task scheduling scheme must carefully handle the \textit{trade-off} between fairness and timely processing of tasks; 
for instance, over-provisioning resources to tasks with enormous demands accelerates their processing; 
however, that may prolong the execution of other tasks as well.  

In this paper, we address the above challenges by designing an online task scheduling scheme for fog computing systems, aiming to achieve ultra-low task latency and multi-resource fairness among tasks. 
The following are our key results and contributions. 
\begin{enumerate}
 \item[$\diamond$] \textbf{System Modeling and Problem Formulation:} To our best knowledge, we are the first to consider multi-resource fairness in the problem of task scheduling for fog computing systems. We develop a system model and formulate the problem under such settings.
 \item[$\diamond$] \textbf{Algorithm Design:} By adopting the idea of dominant resource fairness (DRF) \cite{ghodsi2011dominant} and applying DRL\cite{sutton2018reinforcement}\cite{mao2016resource}, we propose \textit{FairTS}, an efficient online task scheduling scheme that learns directly from experience to achieve ultra-low average task slowdown and ensure the multi-resource fairness among tasks. Notably, FairTS well balances the timely processing of individual tasks and the overall fairness of resource sharing among all tasks. 
 \item[$\diamond$] \textbf{Experimental Verification:} We conduct simulations to evaluate the performance of FairTS. Results show that FairTS effectively shortens task slowdown while inducing better resource fairness than benchmark schemes. 
\end{enumerate}

The rest of the paper is organized as follows. Section II presents the system model and problem formulation. Then, Section III elaborates the design of the task scheduling scheme. 
Section IV presents simulation results, while Section V concludes the paper.

\begin{figure}[!t]
    \setlength{\abovecaptionskip}{-0.0cm} 
    \centering
    \includegraphics[scale=.35]{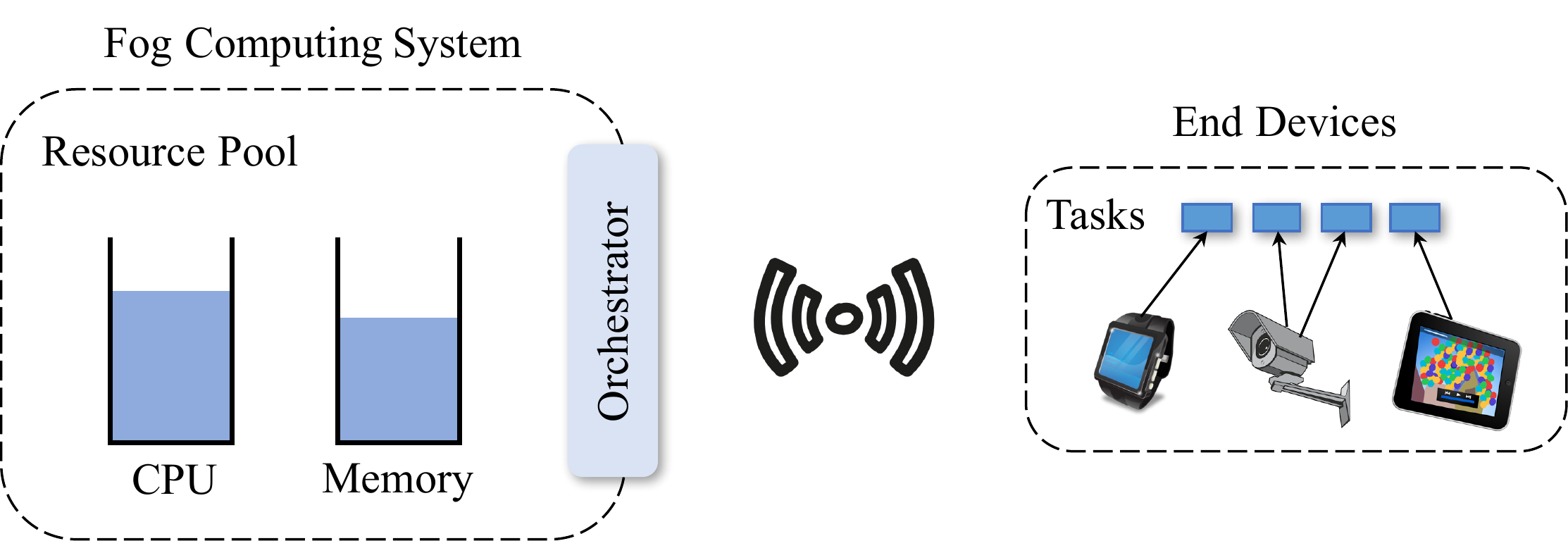}
    \caption{An example of the system model.}
    \label{model}
    \vspace{-0.5cm}
\end{figure}

\section{System Model $\&$ Problem Formulation}
In this section, we present the model and problem formulation for the online task scheduling in fog computing systems. 
\subsection{Basic Model}
We consider a fog computing system which employs a dedicated orchestrator to serve tasks that are generated from end devices on a time-step basis. 
To process these tasks, the orchestrator should allocate resources for each of them upon a resource abstraction layer in a \textit{first-in-first-out} fashion. 
By leveraging state-of-the-art virtualization techniques \cite{kuo2018integrated}, the resource abstraction layer provides various resource pools, each corresponding to one type of computation resource, 
\textit{e.g.}, CPU and memory. 
We show such system model in Figure \ref{model}.

In fog computing system, we assume that the resource abstraction layer consists of resource pools for $m$ types of resources.
The $i$-th resource pool has a capacity of $c_i$, and the resource capacity vector is denoted by $\mathbf{c} = (c_1, c_2, \cdots, c_m)$. 

At the beginning of each time step, end devices generate new tasks awaiting to be uploaded and processed by the fog computing system. 
Each task is assumed \textit{splittable}; \textit{i.e.}, by allocating extra resources to the task, its treatment can be further accelerated proportionally.
For each task $j$, it is characterized by its arriving time $t^{\text{arr}}_j$, starting time $t_{j}$, finishing time $t^{\text{fin}}_{j}$,
bandwidth demand $d_{j, \text{BW}}$ for uploading, and computation demand $\mathbf{d}_j \in \mathbb{Z}^{m}_{+}$ such that $d_{j, i}$ denotes the number of resource units for $i$-th type of resource. 
Note that the starting time $t_j$ and finishing time $t^{\text{fin}}_{j}$ are unknown upon task $j$'s arrival; 
they depend on the orchestrator's scheduling \textit{decisions} 
on \textit{when} to be admitted to serve and \textit{how many} resources the task is allocated with.

Correspondingly, in each time step, the orchestrator conducts task scheduling in two stages. 

In the \textit{first} stage, considering the enormous number of tasks being generated in each time step and the limited wireless capacity, denoted by $c_{\text{BW}}$, between end devices and the fog computing system, 
the orchestrator should decide which subset of tasks to admit and serve.
The starting time $t_j$ for each task $j$ is determined in this stage.

In the \textit{second} stage, the orchestrator allocates available resources to the admitted tasks. 
Particularly, for each task $j$, we assume that the orchestrator allocates it with an integral \textit{multiple} of $\mathbf{d}_{j}$, denoted by $k_j$. 
Such quota-based allocation facilitates more efficient resource management.
Besides, the resource allocation is assumed \textit{non-preemptive}; 
that being said, each task $j$'s allocated resources are exclusive and will not be retrieved by the orchestrator before its completion.
As a result, $k_j$ directly determines task $j$'s execution delay and hence finishing time $t^{\text{fin}}_{j}$.

\subsection{Task Delay Model}
One of the key measures to tasks is the delay between their arrival and completion, especially for tasks with real-time requirements, 
which we refer to as \textit{dwell time}, calculated by $t^{\text{fin}}_{j} - t^{\text{arr}}_{j}$.
Typically, a task's dwell time is divided into the following types of delays.

\textbf{Waiting Delay:} 
The arriving tasks have to stay on end devices before getting admitted and uploaded to the fog computing system. 
For each task $j$, we refer to the period between its arrival time and its starting time (when it is admitted) 
as its \textit{waiting delay}, 
denoted by $t_{j} - t^{\text{arr}}_{j}$.

\textbf{Transmission Delay:} 
Once admitted, the task is transferred to the fog system through the wireless link. 
Considering 
the proximity between fog and end devices, as well as the recent trend in $5$G with ultra-dense deployment of access points and high-speed transmission, 
we assume that task's transmission delay is negligible compared to time step length\footnote{Our model can be easily extended to more general cases with non-zero transmission delays.}. 

\textbf{Execution Delay:} 
For each task $j$, we define its \textit{length} $l_j$ as its execution time if given exactly its demanded resource, \textit{i.e.}, when $k_j = 1$.
Moreover, if allocated with more quotas of resources, \textit{i.e.}, when $k_j > 1$, then its execution can be further accelerated.
However, more resource allocation does not necessarily lead to arbitrary short execution delay, since additional overheads may also be induced, in terms of synchronization and data sharing between sub-tasks.
To take them into account, 
we adopt the recently proposed execution time model \cite{dutton2008parallel}. 
Thus for task $j$, we define its execution delay as a function of $k_j$, \textit{i.e.},
\begin{equation}\label{exe_time}
    t^{\text{exe}}_j = \phi(k_j) \triangleq \left\lceil \frac{l_j}{k_j} + b_j \cdot (k_j-1) \right\rceil,
\end{equation}
where $b_j$ denotes the overhead factor that characterizes the impact of such overheads for task $j$.

Consequently, we can write the finishing time of task $j$ as a function of its starting time $t_j$ and multiple $k_j$
\begin{equation}
	\begin{array}{l}
		t^{\text{fin}}_{j} = \psi(t_j, k_j) = (t_j - t^{\text{arr}}_{j}) + t^{\text{exe}}_{j}.
	\end{array}
\end{equation}

\subsection{Scheduling Objectives}
In the following, we discuss the main objectives to achieve in fog computing systems. Particularly, we consider such objectives over a fixed set of tasks, denoted by $\mathcal{J}$, each of them arriving to the system in an online fashion. 

\textbf{Task Slowdown:} 
When it comes to task scheduling, timely processing is key to delivering best quality-of-service. 
To this end, one can choose to optimize the average task dwell time. However, this may bias the resource allocation towards tasks with lighter computation demand and meanwhile defer the execution of tasks with greater demand.
To address such issues, we normalize tasks' dwell time by their respective lengths. 
Formally, for each task $j$, we consider its \textit{slowdown}, defined as
\begin{equation}\label{slowdown}
    f_j = \frac{t^{\text{fin}}_j - t^{\text{arr}}_j}{l_j}.
\end{equation}
We thus switch to minimize the average task slowdown, \textit{i.e.}, 
\begin{equation}\label{obj_sd}
    \frac{1}{|\mathcal{J}|} \sum_{j \in \mathcal{J}} f_j.
\end{equation}

\textbf{Multi-Resource Fairness:} 
Besides timely processing, it is also necessary to ensure fair resource sharing so that tasks acquire timely processing with resource allocation proportional to their actual demand.
Considering the resource type variety in fog computing systems,  
we adopt dominant resource fairness (DRF) \cite{ghodsi2011dominant}, a recently proposed multi-resource fairness policy. 
The key idea of DRF is to \textit{maximize the smallest dominant share} across tasks,
where for task $j$, its \textit{share} of the $i$-th type resource is $( k_{j} \cdot d_{j, i} / c_{i} )$, and its \textit{dominant resource share} is 
\begin{equation}
    g_j \triangleq \max
    \left\{ \frac{k_j \cdot d_{j, 1}}{c_1}, \cdots, \frac{k_j \cdot d_{j, m}}{c_m}, \frac{d_{j, \text{BW}}}{c_{\text{BW}}} \right\},
\end{equation}
\textit{i.e.}, the maximum among shares of each resource to task $j$.
To ensure balanced dominant resource share across tasks, we aim to reduce the variance across tasks' dominant shares
\begin{equation}\label{obj_fn}
    \frac{1}{|\mathcal{J}|} \sum_{j\in \mathcal{J}} \left( g_j - \bar{g} \right)^2,
\end{equation}
where $\bar{g}$ is the average dominant resource share of tasks.

\subsection{Problem Formulation}
Combining (\ref{obj_sd}) and (\ref{obj_fn}), we target at shortening average task slowdown while guaranteeing a fair resource sharing among tasks. The task scheduling problem is thus defined as follows.
\begin{equation}\label{problem}
\renewcommand{\arraystretch}{1.5}
\begin{array}{lll}
    \underset{ \{t_j, k_j\}_{j} }{\text{Minimize}} ~~~~ & 
    \displaystyle
    \frac{1}{|\mathcal{J}|}\sum_{j \in \mathcal{J}} f_j + \beta \cdot \frac{1}{|\mathcal{J}|} \sum_{j\in \mathcal{J}} \left( g_j - \bar{g} \right)^2 &  \\
    \text{Subject to} ~~~~ & 
    \displaystyle
    t_j \geq t^{\text{arr}}_j\ \text{and}\ k_j \in \mathbb{N}^+ \ \forall\ j \in \mathcal{J}, 
    & (\text{C}1) \\
    & \displaystyle
    \sum_{j\in \mathcal{J}} \mathds{1}_{\{t_j=t\}} d_{j, \text{BW}} \leq c_{\text{BW}} 
    \ \ \forall\ t, 
    & (\text{C}2)
    \\
    & \displaystyle
    \sum_{j\in \mathcal{J}} \mathds{1}_{\{t \in [t_j, t^{\text{fin}}_j]\}} k_j d_{j, i} \leq c_{i}, \forall t \ \text{and}\ i.
    & (\text{C}3)
\end{array}
\end{equation}
where 
$\beta > 0$ is a constant parameter that weights the importances of shortening task slowdown and resource fairness. 
Besides, $\mathds{1}_{\{t_j=t\}}$ and $\mathds{1}_{\{t \in [t_j, t^{\text{fin}}_j]\}}$ indicate, in time step $t$, whether task $j$ is admitted and whether it is using resource $i$, respectively.
As for the constraints:
(C1) requires that each task starts after its arrival and receives resources as a integral multiple of its demand.
(C2) ensures that the total bandwidth demand of admitted tasks in each time step would not exceed the wireless bandwidth. 
(C3) guarantees total resource consumption in any time step does not exceed its corresponding resource capacity.

\section{Algorithm Design}
With full priori information about task arrivals and their resource demand, 
one can solve Problem (\ref{problem}) optimally, al-though it requires considerably high complexity \cite{brucker1999resource} even for its special case where $k_j=1$ for any task $j$. 
However, usually such information is usually unavailable in practice. 
Inspired by recent success in applying deep reinforcement learning (DRL) techniques to resource management problems \cite{mao2016resource},
we realign our goal to design an online task scheduling scheme that learns from experience to achieve ultra-low task delay with multi-resource fairness.

\begin{figure*}[!t]
    \setlength{\abovecaptionskip}{-0.0cm} 
    \centering
    \subfigure[The state representation at the beginning of time step $t$.]{
        \includegraphics[scale=.34]{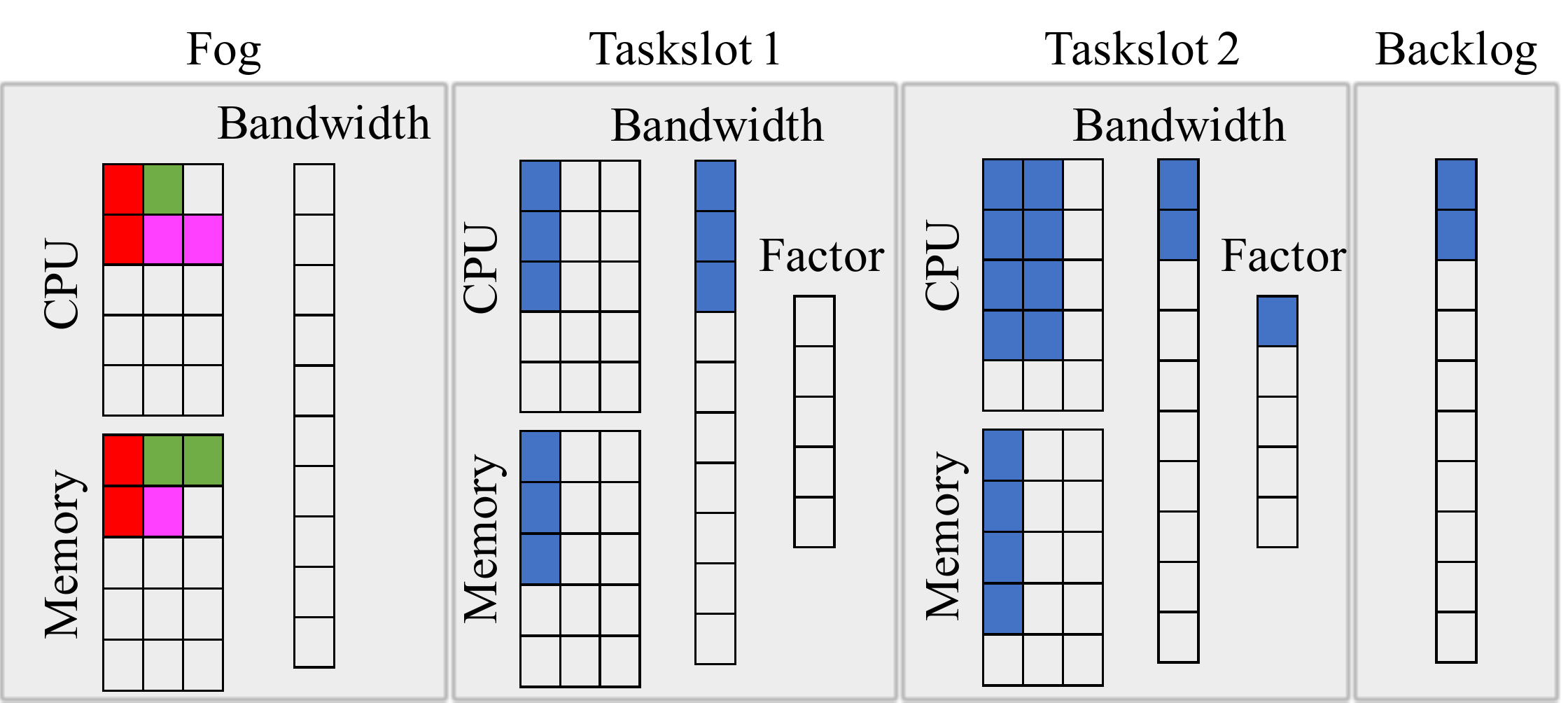}
    }
    \subfigure[The state representation after taking action $(1, 2)$.]{
        \includegraphics[scale=.34]{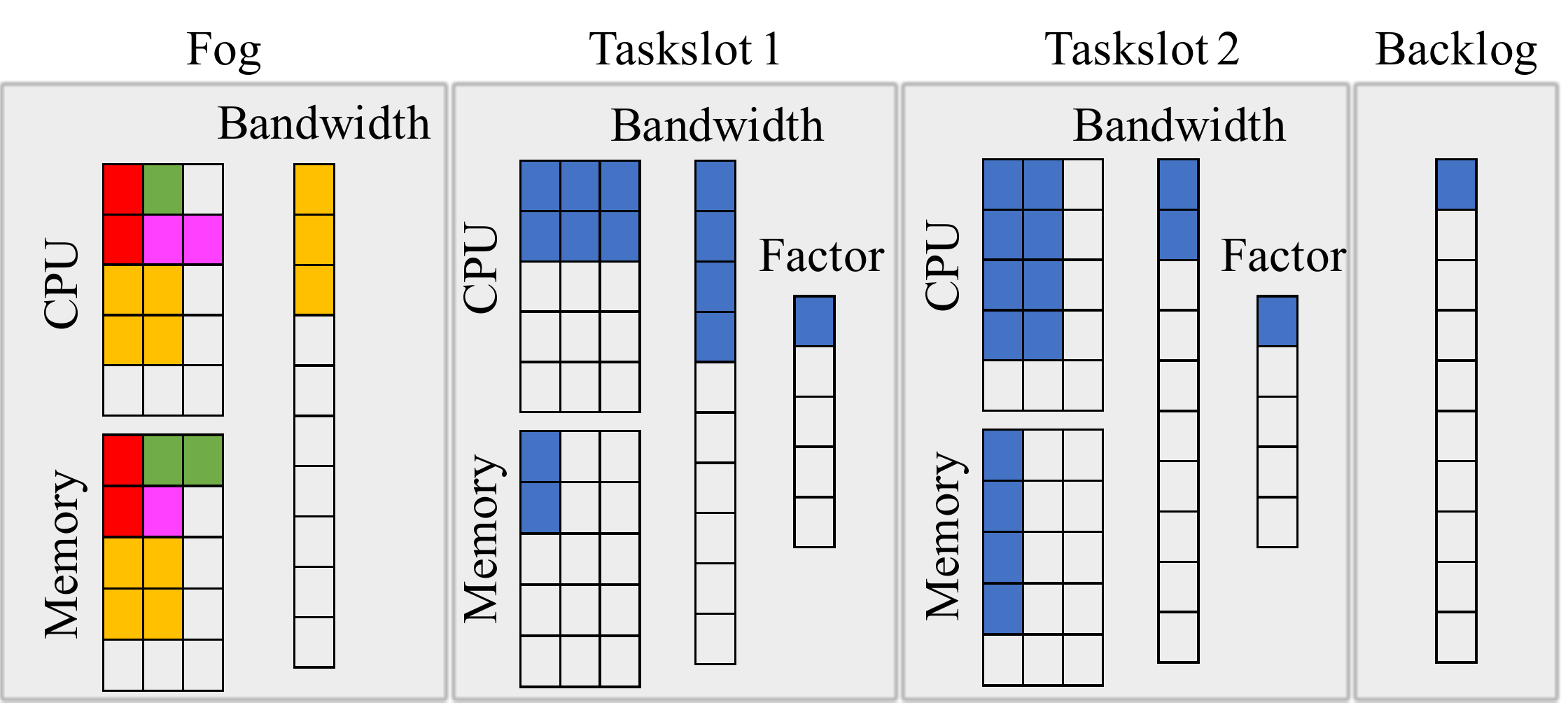}
    }
    \caption{
    	System state evolution before and after taking action $(1, 2)$.
    	\textit{Basic Settings}: We consider two types of resources CPU and memory, both of capacity $3$ units; meanwhile, the wireless link has a capacity of $10$. The orchestrator decides the resource allocation in the next $5$ time steps. There are two Taskslots and the Backlog capacity is $10$.
		State (a): The red task occupies one CPU and one memory unit, and will be finished in two time steps. The green task occupies $1$ CPU, $2$ memory, and its residual execution time is one time step. The purple task will start executing in the next time step, occupying 2 CPU units and one memory unit for one time step. 
       In Taskslot 1, there is a task that demands $1$ CPU unit, $1$ memory unit, and $3$ bandwidth units. Its length is $3$ time steps, and overhead factor is zero. In Taskslot 2, the task demands $2$ CPU units, $1$ memory unit, and $2$ bandwidth units. Its length is $4$ time steps and overhead factor is $1$. 
       There are two tasks in the Backlog, but no detailed information of those tasks is included. 
       State (b): By taking action $(1, 2)$, the task in Taskslot 1 is chosen, and the allocated resources are $2\times 1$ CPU units, $2 \times 1$ memory units and $3$ bandwidth units. The execution delay will be $\left \lceil \frac{3}{2} + 0\times (2 - 1) \right \rceil = 2$ time steps. The chosen task will be moved to the resource availability image and identified using orange color. Meanwhile, the task on the top of the backlog will be assigned in Taskslot 1 for substitution. 
    }
    \label{state}
\end{figure*}

\subsection{Deep Reinforcement Learning}
A reinforcement learning (RL) problem generally considers a situation in which an online sequential decision making process between an agent and its environment on a time-step basis. In each time step $t$, the agent observes the latest environment state $s_t$ and decides to exert some action $a_t$ to the environment.
Accordingly, the environment evolves to a new state $s_{t+1}$ according to some Markov Decision Process and reveals a reward signal $r_{t}$ back to the agent. 
The agent collects the reward and improves its decision making policy $\boldsymbol{\pi}$, so as to maximize its expected cumulative rewards in the long run, denoted by $\mathbb{E}[\sum_{t=0}^{\infty} \gamma^{t} r_t]$, where $\gamma \in (0, 1]$ is a constant factor that discounts the impact of future rewards.

To solve RL problems with enormous state space size, DRL techniques, 
\textit{e.g.}, \textit{policy gradient} (PG) methods \cite{sutton2018reinforcement}, have enjoyed wide adoption in recent years.
Typically, PG views the agent's policy as a non-linear function $\pi_{\theta}(\cdot)$ with unknown parameters $\theta$, which takes current state as input and outputs the action selection probability vector.
In the following subsections, we first reformulate Problem (\ref{problem}) as an RL problem and then employ PG to solve it.

\subsection{Problem Reformulation}
To reformulate Problem (\ref{problem}), we regard the orchestrator as the agent, and the fog computing system as its environment. 
In each time step, the agent observes the resource availability and resource demand of impending tasks (state), then applies its policy to \textit{repeatedly} pick a task and decide its resource allocation (action) until no more tasks are selected.
In the end, a reward signal is fed back to the agent. 
We elaborate the design of state space, action space, and reward, as follows.

\textbf{State Space:} 
We denote the system state representation between successive actions by two types of images, the image of resource availability and impending task resource demand, respectively. 
Ideally, revealing all images in the current time step conduces to the best possible decision making. 
However, in practice, given numerous tasks arriving in each time step, such information can be overwhelming to maintain.

To address such issues, we design a state representation with a constant size.
In particular, we denote the resource availability image by \textit{Fog}, 
and distinguish two types of image for impending task resource demand by \textit{Taskslot} and \textit{Backlog}, respectively.
Fog shows the resource allocation in the next few time steps.
A Taskslot records the resource demand $\mathbf{d}_{j}$, bandwidth demand $d_{BW}$, overhead factor $b_j$, and the length $l_j$, of a particular task $j$. 
The system state only includes the Taskslots of the first $n$ impending tasks.
The rest of impending tasks is concluded in the Backlog image without detailed features, and we use $c_{\text{BL}}$ to denote the capacity of Backlog.
We show an example of system state representation in Figure \ref{state} with $n=2$ and $c_{\text{BL}}=10$.

\textbf{Action Space:} During each time step $t$, the orchestrator should choose a subset of the $n$ tasks from Taskslots and decide corresponding resource allocation, with an action space size of $O(2^n)$.
To conduct more timely and efficient scheduling decisions, 
we redefine the action as picking one of the impending tasks and deciding its resource allocation.
Particularly, each action $a$ is denoted by $(j, k_j)$, 
where $j$ is the Taskslot ID to be chosen ($0$ for void action) and $k_j$ indicates that task $j$ is allocated with $k_j$ multiple of its resource demand. 
For example, action $(2, 3)$ means ``picking the task in Taskslot $2$ and allocate $3$ times of resources that it demands".  
In each time step, we allow the agent to take more than one action in each time step.

After performing a valid action, the system state transits to a new state, where the chosen task $j$'s Taskslot image is cleared out, 
and the resource  will be updated according to $t_j$ and $k_j$. 
Next, the Taskslot image of the task on the top of the backlog (if exists) will be included into the new state. 

Note that an action may be invalid by the time it is chosen.
For instance, an action with selected task with length $W$ is invalid, 
if the available resources in Fog can not satisfy the allocation during any time step between $[t, t+W)$.
Whenever an invalid or void action is taken, the orchestrator stops the scheduling process and move forwards to the next time step. 

\textbf{Rewards:} 
The reward signal should be designed to align the agent's goal of maximizing cumulative reward to minimizing average slowdown with multi-resource fairness guarantee, as in (\ref{problem}).
To this end, we define the reward at the end of each time step $t$ as
\begin{equation}\label{reward}
    r_t \triangleq - \left[ \sum_{j\in \mathcal{J}_t} \frac{1}{l_j} + \beta \cdot \frac{1}{\left \vert \mathcal{J}^{\text{ing}}_t \right \vert} \sum_{j\in \mathcal{J}^{\text{ing}}_t} (g_j - \bar{g})^2 \right].
\end{equation}
where $\mathcal{J}_t$ denotes the set of tasks dwelling in the system in time $t$, and $\mathcal{J}^{\text{ing}}_t \subseteq \mathcal{J}_{t} $ as the set of tasks being run. Note that reward (\ref{reward}) is revealed only at the end of each time step, \textit{i.e.}, when the action is void or infeasible. Other intermedia actions receives no reward. 

%

\begin{algorithm}[!b]
    \caption{Fair Task Scheduling (FairTS) scheme}
    \begin{algorithmic}[1]
    \label{alg}
    \FOR {each iteration}
        \STATE $\Delta \theta \leftarrow 0$
        \FOR {each episode $i$}
            \STATE Sample a trajectory $s^i_1, a^i_1, r^i_1, \cdots, s^i_{L_i}, a^i_{L_i}, r^i_{L_i}$
            \STATE Compute returns $v^i_t = \sum_{s=t}^{L_i} \gamma^{s-t} r^i_s$
        \ENDFOR
        \STATE $L \leftarrow \min_{i} L_i$
        \FOR {$t=1$ to $L$}
            \STATE Calculate baseline $b_t = \frac{1}{N} \sum_{i=1}^N v^i_{t}$
            \FOR {$i=1$ to $N$}
                \STATE $\Delta \theta \leftarrow \Delta \theta + \alpha \cdot \nabla_\theta \log \pi_{\theta}(s^i_t, a^i_t) \cdot (v^i_t - b_t)$
            \ENDFOR
        \ENDFOR
        \STATE $\theta \leftarrow \theta + \Delta \theta$
    \ENDFOR 
    \end{algorithmic}
\end{algorithm}

\subsection{Training Algorithm} 
We design a Fair Task Scheduling (FairTS) scheme, and the pseudocode is shown in Algorithm \ref{alg}.
We construct the policy $\pi_{\theta}(\cdot)$ as a forward neural network with two fully connected layers with its weight parameters $\theta$ to characterize the policy. 
Then we train the policy using policy gradient method over a synthesis dataset of a fixed set of tasks on an iteration basis, as shown in Algorithm \ref{alg}.
During each iteration, we simulate $N$ episodes and record their trajectories, 
where each episode is simulated based on the current policy and terminates when all tasks are completed.
At the end of each iteration, we leverage such trajectories to improve the policy.

\section{Simulation Results}

\subsection{Basic Settings}

\textbf{Workload:} We evaluate FairTS using the settings in Table \ref{settings}. 
Tasks arrive online in a poisson process with mean rate $0.8$. 
We assume that all tasks have the same bandwidth demand $1$, and the same overhead factor $2$. 
According to the length and size of resource demand, we classify tasks into two categories: elephant or mice. 
An elephant task has a length of $5$ time steps, and its resource demand is randomly distributed within the range of $80\%$ to $100\%$ of the resource capacity. 
A mice task's length is $1$, and the resource demand is sampled from $10\%$ to $20\%$ of resource capacity. A task is equally likely to be elephant or mice.  

\textbf{Benchmark:} We compare FairTS with two state-of-the-art benchmarks. 
For each non-void action, the Taskslot is chosen uniformly over the $n$ Taskslots at random and the number of resources is decided either (1) randomly (\textbf{Random}) or (2) by Shortest Execution Time (\textbf{SET}) \cite{dutton2008parallel}. When the Taskslots are empty, both benchmarks return void action. 

\begin{table}[!tbp]
\begin{center}
    \begin{tabular}{|c|c|}
    \hline
    \textbf{Parameters} & \textbf{Value} \\
    \hline
    Number of CPU units $c_1$ & $5$ \\
    Number of memory units $c_2$ & $10$ \\
    Number of bandwidth units $c_{\text{BW}}$ & $10$\\
    \hline
    Learning rate $\alpha$ & $10^{-5}$ \\
    Weight parameter $\beta$ & $0, 200, 400, 500$ \\
    Discounting factor $\gamma$ & $1$ \\
    \hline
    Lookahead window size $W$ & $5$ \\
    Number of Taskslots $n$ & $2, 3, 4, 5$\\
    Backlog capacity $c_{\text{BL}}$ & $100$ \\
    \hline
    \end{tabular}
\end{center}
\caption{Simulation settings.}
\label{settings}
\vspace{-0.5cm}
\end{table}

\begin{figure}
    \setlength{\abovecaptionskip}{-0.0cm} 
    \centering
    \subfigure[Average task slowdown.]{
        \includegraphics[scale=.15]{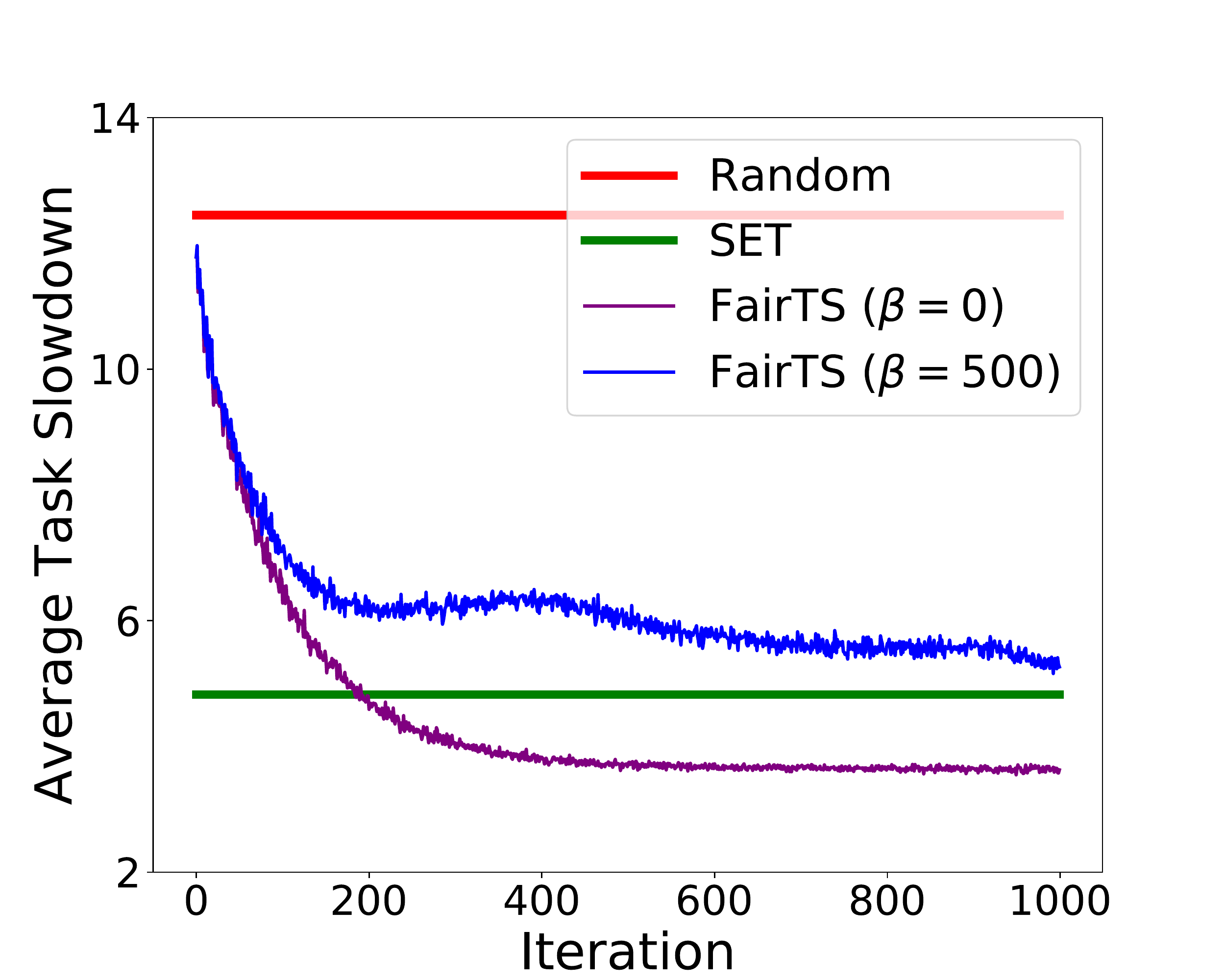}
    }
    \subfigure[Dominant share variance.]{
        \includegraphics[scale=.15]{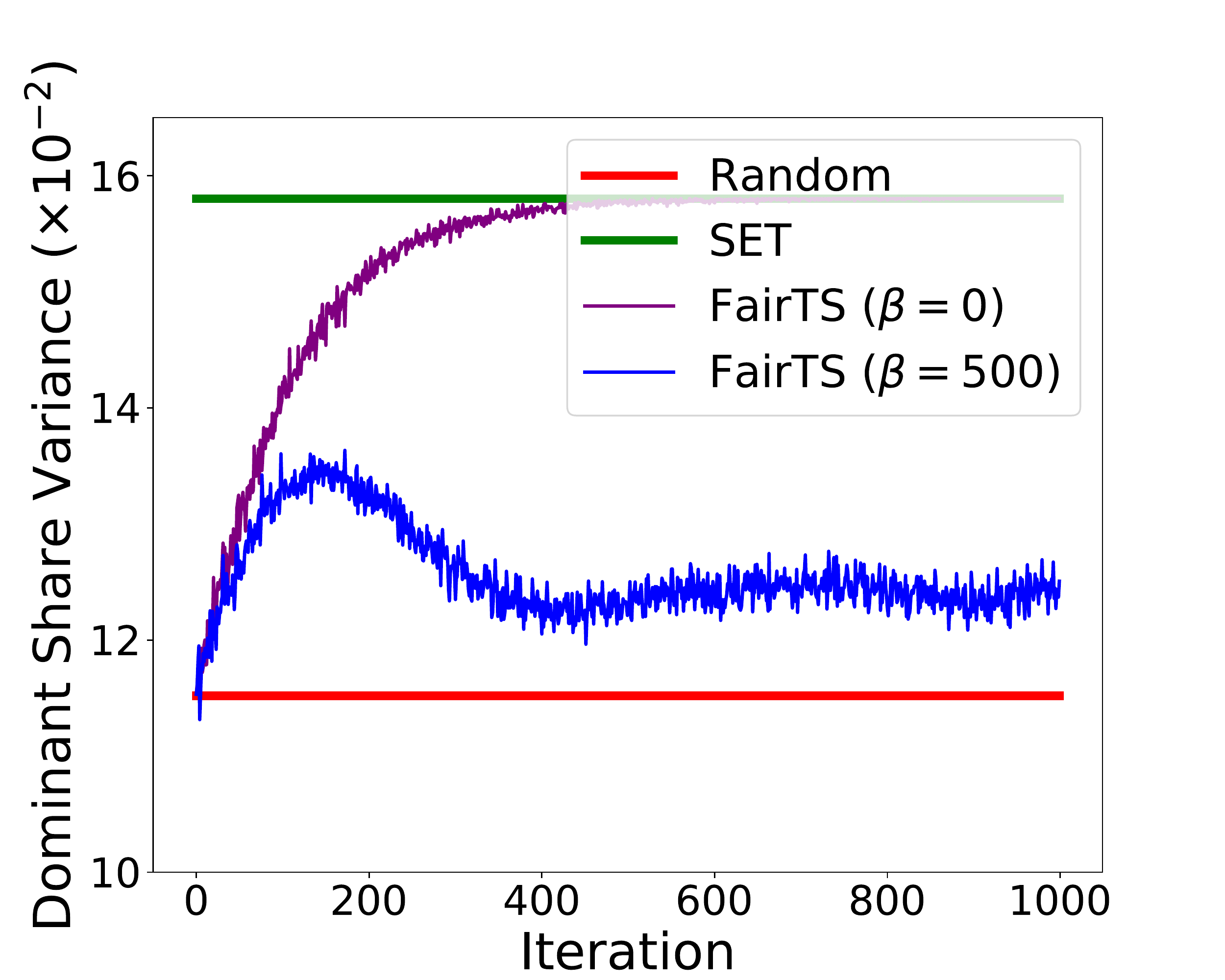}
    }
    \caption{Comparison of FairTS and heuristics in terms of (a) average task slowdown and (b) dominant share variance.}
    \label{curve}
    \vspace{-0.5cm}
\end{figure}

\subsection{Evaluation Results}
Unless specified, the number of Taskslots is default as $3$.
Figure \ref{curve} presents the average task slowdown and dominant share variance of Random, SET and FairTS (with $\beta=0$ and $\beta=500$) over iterations.
Each data point is an average of $100$ runs. We make the following observations. 
\begin{itemize}
    \item[$\diamond$] Random incurs extremely high average task slowdown, because all decisions are taken randomly without utilizing any system information.
    \item[$\diamond$] Random leads to small dominant share variance, which means that the dominant shares of most tasks are at the same level, implying a fair resource allocation.
    \item[$\diamond$] SET has relatively low slowdown but large dominant share variance, implying unfair resource allocation. 
    \item[$\diamond$] The average task slowdown of FairTS decreases with iterations, while the dominant share variance increases. 
    \item[$\diamond$] Increasing $\beta$ from $0$ to $500$ will result in an increase in the average task slowdown and a reduction in the dominant share variance.
\end{itemize}

\begin{figure}
    \setlength{\abovecaptionskip}{-0.0cm} 
    \centering
    \subfigure[Average task slowdown.]{
        \includegraphics[scale=.15]{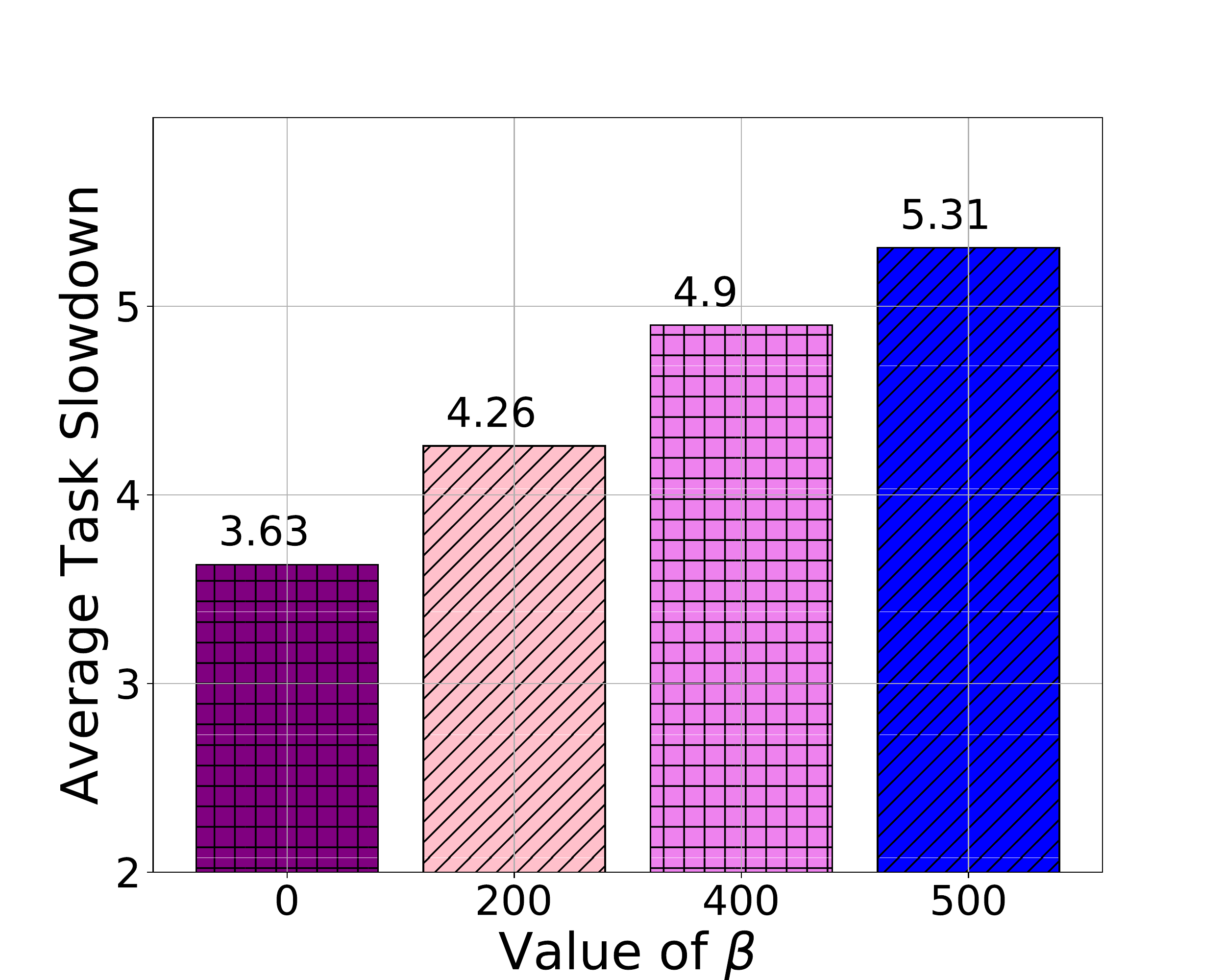}
    }
    \subfigure[Dominant share variance.]{
        \includegraphics[scale=.15]{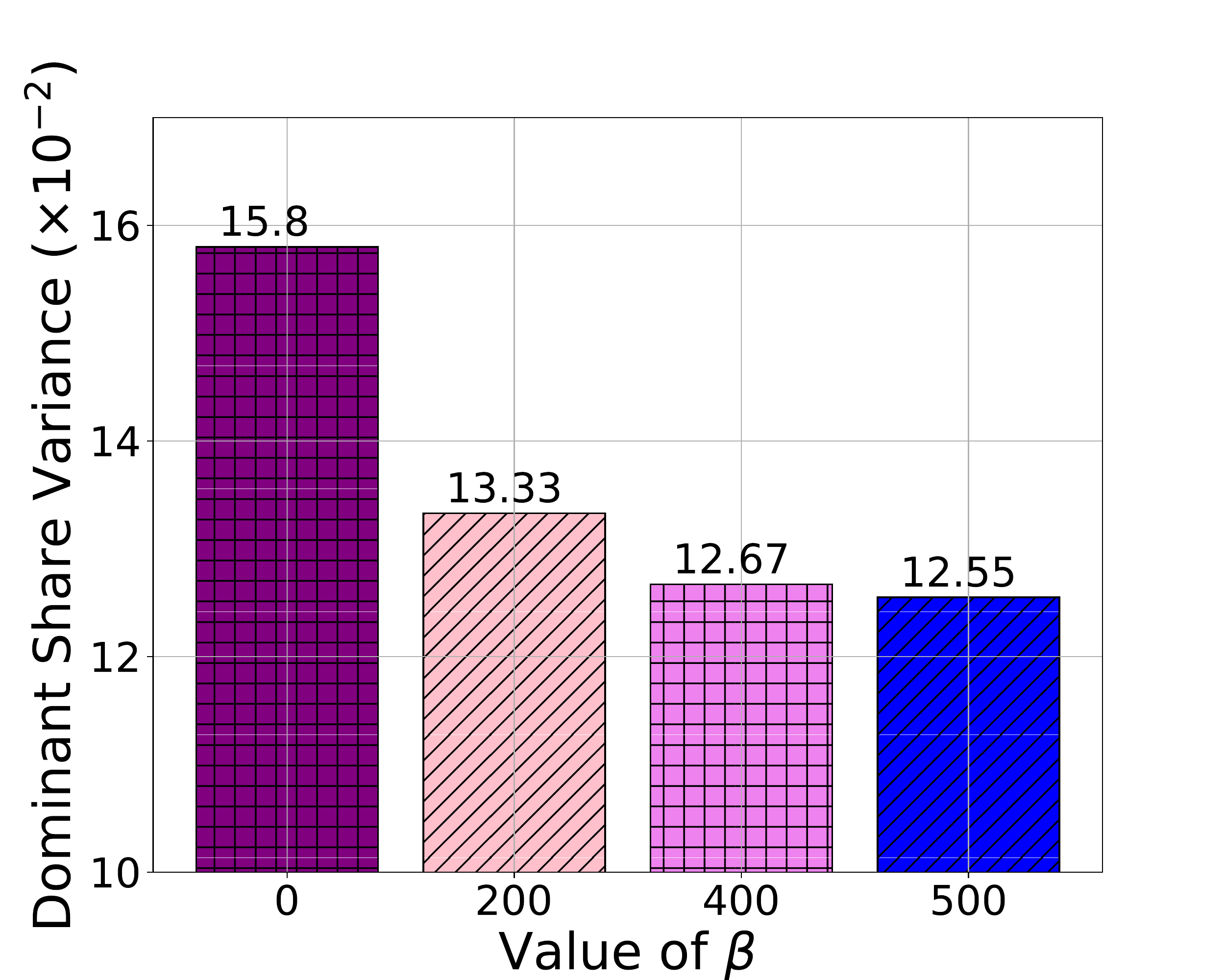}
    }
    \caption{Performance of FairTS with different setting of parameter $\beta$.}
    \label{hist_beta}
    \vspace{-0.5cm}
\end{figure}

Next, we investigate the effect of weighting parameter $\beta$ in detail. From Problem (\ref{problem}), we know that with larger $\beta$, more concentration is addressed on resource fairness. Figure \ref{hist_beta} shows the simulation results under different settings of $\beta$. As $\beta$ increases from $0$ to $500$, the average task slowdown increases, and the dominant share variance decreases gradually. We conclude that larger $\beta$ incurs higher average task slowdown, but more fair resource allocation.

\begin{figure}
    \setlength{\abovecaptionskip}{-0.0cm} 
    \centering
    \subfigure[Average task slowdown.]{
        \includegraphics[scale=.15]{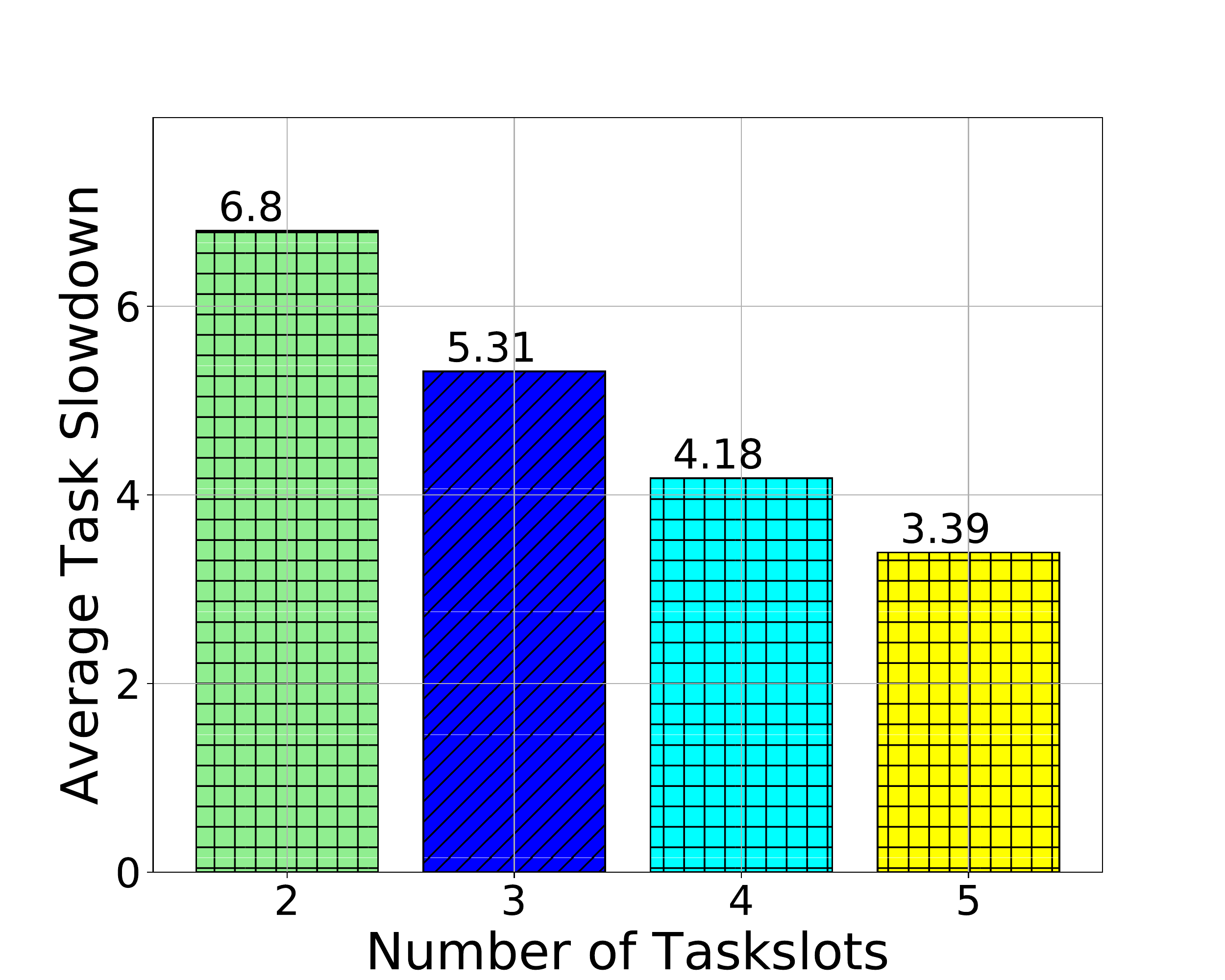}
    }
    \subfigure[Dominant share variance.]{
        \includegraphics[scale=.15]{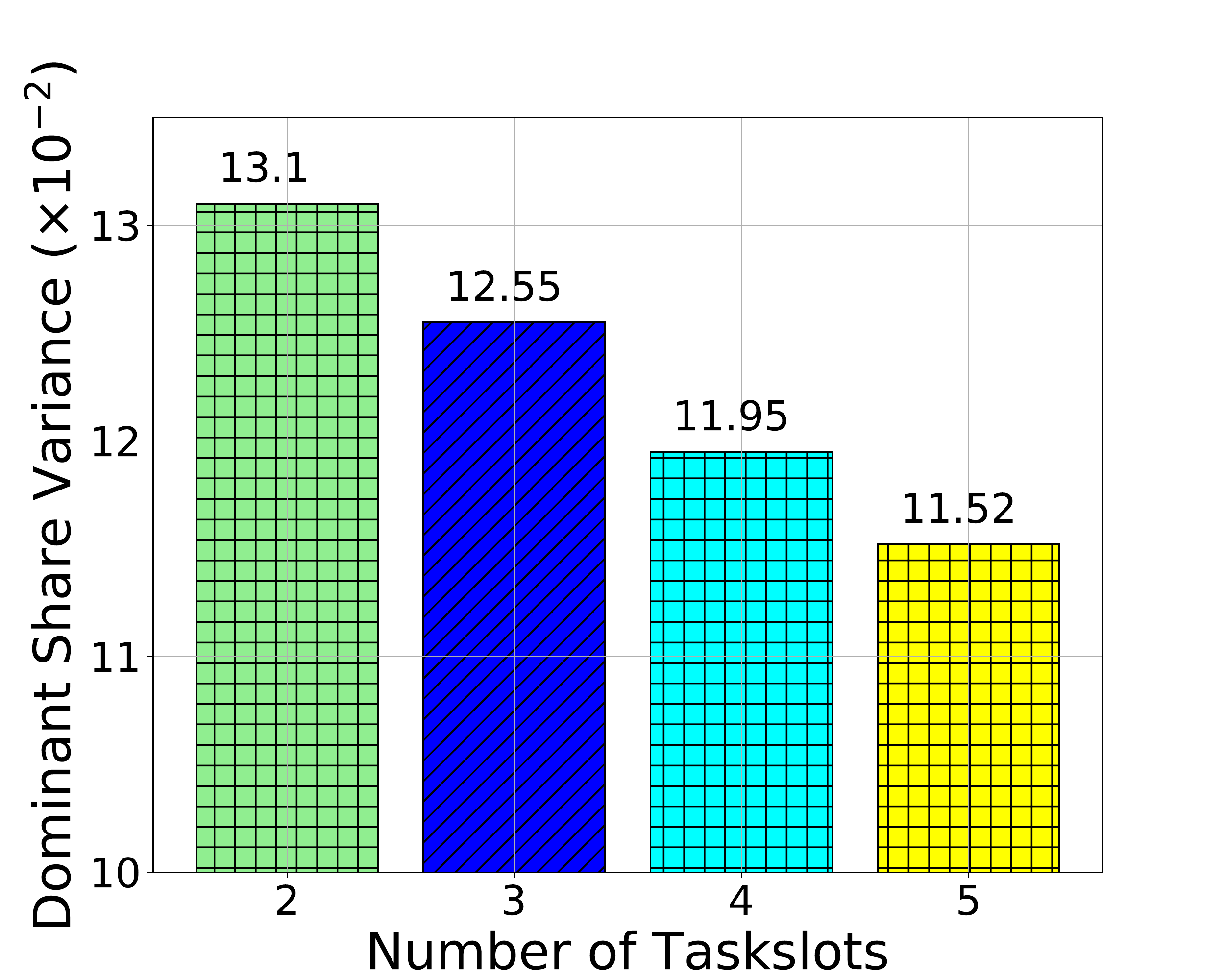}
    }
    \caption{Performance of FairTS with different setting of Taskslot number $n$.}
    \label{hist_n}
    \vspace{-0.5cm}
\end{figure}

In Figure \ref{hist_n}, we present the simulation results under different settings of Taskslot number $n$. 
As expected, we observe that more Taskslots brings better performance. 
However, increasing the number of Taskslots will enlarge the state size significantly, which brings higher training complexity. 


\section{Conclusion}
In this paper, we studied the online task scheduling problem for fog computing systems, considering both task slowdown and multi-resource fairness. 
By applying DRF and DRL, we proposed \textit{FairTS}, an efficient online fair task scheduling scheme. Evaluation results showed that FairTS outperforms the state-of-the-art heuristics.

\bibliography{references}

\end{document}